\documentclass[lettersize,journal]{IEEEtran}
\usepackage{amsmath,amsfonts}
\usepackage{algorithmic}
\usepackage{algorithm}
\usepackage{array}
\usepackage[caption=false,font=normalsize,labelfont=sf,textfont=sf]{subfig}
\usepackage{textcomp}
\usepackage{stfloats}
\usepackage{url}
\usepackage{verbatim}
\usepackage{graphicx}

\usepackage{cite}
\usepackage{amsmath} 
\usepackage{color}
\usepackage[table,xcdraw]{xcolor}
\usepackage{amssymb}

\usepackage{hyperref}
\usepackage{multirow}
\usepackage{booktabs}

\begin{document}

\title{Fragment-Masked Diffusion \\ for Molecular Optimization}

\author{
Kun Li$^{1}$, Xiantao Cai$^{1}$, Jia Wu$^{2}$, Shirui Pan$^{3}$,  Huiting Xu$^{4}$, Bo Du$^{1}$, Wenbin Hu$^{1,*}$  \thanks{*Correspoonding Author.}\\
$^{1}$School of Computer Science, Wuhan University, Wuhan, Hubei, China\\
$^{2}$Department of Computing, Macquarie University, Sydney, NSW, Australia\\
$^{3}$School of Information and Communication Technology, Griffith University, Brisbane, Queensland, Australia\\
$^{4}$Department of Abdominal Oncology, Hubei Cancer Hospital, Wuhan, Hubei, China\\
\{likun98, caixiantao, dubo, hwb\}@whu.edu.cn, Jia.wu@mq.edu.au, s.pan@griffith.edu.au, annexu333@126.com
}



\markboth{}%
{Shell \MakeLowercase{\textit{et al.}}: A Sample Article Using IEEEtran.cls for IEEE Journals}


\maketitle

\begin{abstract}
Molecular optimization is a crucial aspect of drug discovery, aimed at refining molecular structures to enhance drug efficacy and minimize side effects, ultimately accelerating the overall drug development process. Many molecular optimization methods have been proposed, significantly advancing drug discovery. These methods primarily on understanding the specific drug target structures or their hypothesized roles in combating diseases. However, challenges such as a limited number of available targets and a difficulty capturing clear structures hinder innovative drug development. In contrast, phenotypic drug discovery (PDD) does not depend on clear target structures and can identify hits with novel and unbiased polypharmacology signatures. As a result, PDD-based molecular optimization can reduce potential safety risks while optimizing phenotypic activity, thereby increasing the likelihood of clinical success. Therefore, we propose a fragment-masked molecular optimization method based on PDD (\textbf{FMOP}). FMOP employs a regression-free diffusion model to conditionally optimize the molecular masked regions, effectively generating new molecules with similar scaffolds. On the large-scale drug response dataset GDSCv2, we optimize the potential molecules across all 985 cell lines. The overall experiments demonstrate that the in-silico optimization success rate reaches 95.4\%, with an average efficacy increase of 7.5\%. Additionally, we conduct extensive ablation and visualization experiments, confirming that FMOP is an effective and robust molecular optimization method. The code is available at: \url{https://anonymous.4open.science/r/FMOP-98C2}. 
\end{abstract}

\begin{IEEEkeywords}
Molecular optimization, fragment-masked, diffusion model, phenotypic drug discovery, drug discovery.
\end{IEEEkeywords}

\section{Introduction}

\IEEEPARstart{M}{olecular} optimization plays a crucial role in drug discovery \cite{Opt_review}, which involves the modification and improvement of lead compounds identified through initial screening to enhance their drug-like properties \cite{mol_property_1}. Historically, molecular optimization is planned manually according to knowledge and experience in the pharmacology, and optimized through fragment-based screening or synthesis \cite{fragOpt}. However, manual molecular optimization is not easily scalable to different needs and cannot be automated for large-scale optimization; thus, this strategy is insufficient for meeting the demands of current drug discovery \cite{RethinkDD}.

In recent years, deep learning (DL) methods, particularly diffusion models \cite{DDPM}, have been observed to effectively optimize molecules that meet specific conditions \cite{DecompOpt,gu2024aligning}, with the potential to accelerate traditional paradigms. Molecular optimization can rapidly identify potential drug candidates using existing experimental data and molecular structures, reducing the need for blind experimentation and enhancing research efficiency \cite{choudhuri2023recent}.




\begin{figure}[t!]
    \centering
    \includegraphics[width=0.99\linewidth]{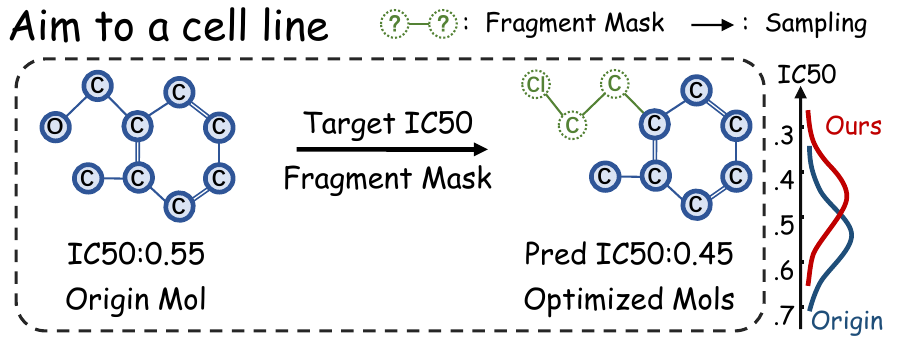}
    \caption{PDD molecular optimization task. The diagram on the right compares the IC50 distributions of original and optimized molecules obtained by our method. }
    \label{fig:motivation}
\end{figure}

Several molecular optimization methods have been proposed to enhance candidate molecule performance, such as target pocket- \cite{OptDTI} and property-based molecular optimization \cite{Prompt-MolOpt, OptProp3, DST, MARS}. Target pocket-based molecular optimization methods rely on understanding specific pocket structures and their hypothesized roles in combating diseases \cite{adv:pdd2}. Challenges such as a limited number of available targets and difficulty capturing clear structures hinder innovative drug development. In contrast, phenotypic drug discovery (PDD) \cite{adv:pdd1} does not depend on well-defined target structures and can identify hit compounds with novel and multi-target properties. PDD emphasizes the phenotypic effects of molecules within disease-related biological systems \cite{Nature_DD} and has significantly contributed to the discovery of first-in-class drugs \cite{dic:pdd}. By observing the phenotypic changes that molecules induce in cells, tissues, or organisms, PDD identifies potential drugs without requiring prior knowledge about specific targets \cite{pdd3}. As a result, PDD's potential as a drug discovery tool that addressing complex diseases that are not yet fully understood by the scientific community is evident. Extensive PDD research has been conducted, with significant efforts made in constructing relevant datasets such as the genomics of drug sensitivity in cancer (GDSC) \cite{GDSC,GDSCv1}. Based on these datasets, numerous artificial intelligence-driven methods \cite{drp,CLDR,MSDA} have been proposed to predict drug responses against specific cell lines, typically quantified by the half-maximal inhibitory concentration (IC50) \cite{numberisbetter}. These advancements have significantly accelerated PDD research \cite{futurePDD}.

Based on prior knowledge, molecular optimization methods specifically designed for PDD have not been proposed, primarily due to several challenges in this field. First, encoding molecular interactions with cellular systems differs significantly from existing target- and properties-based molecular optimization methods. Second, the evaluation metrics for these tasks vary, making direct adaptation challenging. While the physicochemical drug properties (e.g., lipophilicity and solubility) can be measured quickly and cost-effectively \cite{OptProp}, and drug-target affinities can be accurately predicted using existing virtual screening techniques \cite{dta3,dti2}, these methods are not directly applicable to PDD. In the context of PDD optimization, the ideal evaluation metric is the IC50 value, which reflects the interaction between molecules and cell lines. Although determining IC50 values experimentally is time-consuming and expensive, they can be predicted by drug response prediction (DRP) models within a certain margin of error. By evaluating the predicted IC50 distribution of an optimized molecular set, we can assess the optimization model's effectiveness, thereby mitigating the impact of prediction errors.


For this reason, we propose a \textbf{f}ragment-\textbf{m}asked molecular \textbf{op}timization method for the PDD (\textbf{FMOP}). As shown is Fig. \ref{fig:motivation}, the FMOP method's optimization conditions include the initial molecule to be optimized, its corresponding masked fragments, the target cell line, and the IC50 value. Notably, the IC50 is an efficacy measure of a single drug response against one specific cell line, obtained through the wet experiment. The output is a batch of optimized molecules under specified conditions. FMOP employs a regression-free diffusion model to conditionally optimize the molecules' masked regions, effectively generating new molecules with similar scaffolds and improving IC50. Specifically, molecules' fragment masks are primarily based on scaffolds and side chains in the molecular structure. First, we apply rule-based constraints to the fragmentation results. Then, we use a pre-trained diffusion model as the generative prior and adjust the unmasked region sampling process during the reverse diffusion iteration using the given drug information, without modifying or conditioning the original diffusion model. 


On the large-scale drug response dataset GDSCv2 \cite{GDSC}, we conducted optimization experiments on all 985 cell lines, demonstrating an in-silico optimization success rate of 95.4\% and an average efficacy increase of 7.5\%. \textit{It is important to emphasize that the optimization task across 985 cell lines is analogous to optimization across different properties. Our method requires training only once to cover all tasks, whereas other methods would need to train independently for each of the 985 cell lines}. Additionally, through extensive ablation and visualization experiments, we further demonstrate that FMOP is an effective and robust molecular optimization method with broad application prospects in PDD. This paper's contributions are as follows:

\begin{enumerate}
    \item We introduce FMOP, a novel fragment-masked molecular optimization method. The FMOP method integrates scaffold-based fragments with rule-based constraints and leverages a pre-trained diffusion model to optimize masked regions according to molecule information and the PDD task conditions, without requiring model training. To the best of our knowledge, the FMOP is the first optimization method for the PDD task.
   
    \item Optimization experiments were conducted across all 985 cell lines in the GDSCv2 dataset, demonstrating a 95.4\% success rate and a 7.5\% average increase in efficacy through optimization. Extensive visualization evaluations further indicate FMOP's robustness and broad applicability.
\end{enumerate}

\section{Related Work}

Molecular optimization aims to improve drug properties, including physicochemical (e.g., solubility, stability, and absorption) and biomedical attributes (e.g., toxicity, target affinity, and drug-cell response) \cite{tap, DropConn, GIaNt}, thereby enhancing drug candidate effectiveness. Existing molecular optimization methods could be broadly categorized into rule- and deep learning-based methods. Rule-based methods \cite{HN-GFN, MARS}, such as pharmacophore modeling and fragment libraries built using JT-VAE \cite{JT-VAE}, relied on predefined structural rules to suggest modifications. In comparison, deep learning-based methods \cite{GeoLDM, GDSS}, such as those utilizing the denoising diffusion probabilistic model (DDPM) \cite{DiffDec,Pocket2mol,DDPM}, demonstrated great potential for molecular optimization. More recent methods were flow-based \cite{FFLOM}, graph-based \cite{HN-GFN, CDGS, MOOD, DiGress, DST} and Transformer-based \cite{Prompt-MolOpt}, integrating diffusion models to enhance optimization outcomes \cite{Graphusion}. These methods integrated DDPM’s ability to generate high-quality and diverse molecules with specific optimization goals and conditions to efficiently refine molecular structures and properties. 

Despite significant progress, current molecular optimization techniques encountered limitations in meeting PDD requirements. Numerous studies have been proposed optimizing physicochemical properties \cite{OptProp2, KG-MTL, DGDFS}, such as toxicity and target affinity \cite{OptDTI,Opt_DTI_2}, demonstrating significant potential in accelerating molecular optimization. However, PDD molecular optimization methods have not been thoroughly investigated. Some drugs showed favorable solubility and stability in vitro but failed to deliver the expected efficacy in practical applications \cite{vitro}. This gap occurred due to current methods disregarding the complexity of cellular environments. In addition, current optimization methods were often limited to specific masks or predicting functional group placements \cite{fragment1, DST, HN-GFN}. These fragment-based methods regarded molecular optimization as the addition or deletion of individual atoms or chemical bonds \cite{OptDTI, fragment2, FrequentSubstructure}. Consequently, when tackling novel mask types or tasks, these methods may struggle to handle complex masking scenarios due to limited training data, thereby restricting their applicability \cite{broker2022fragment}. Most importantly, the training and optimization mechanisms employed by these methods were originally designed to train only on a set of molecules with a specific property. When the properties change, these models have to be retrained, so the requirements of modeling multiple properties at the same time and optimizing the molecule for different properties could not be met if secondary training was required.

\begin{figure*}[t!]
    \centering
    \includegraphics[width=0.98\linewidth]{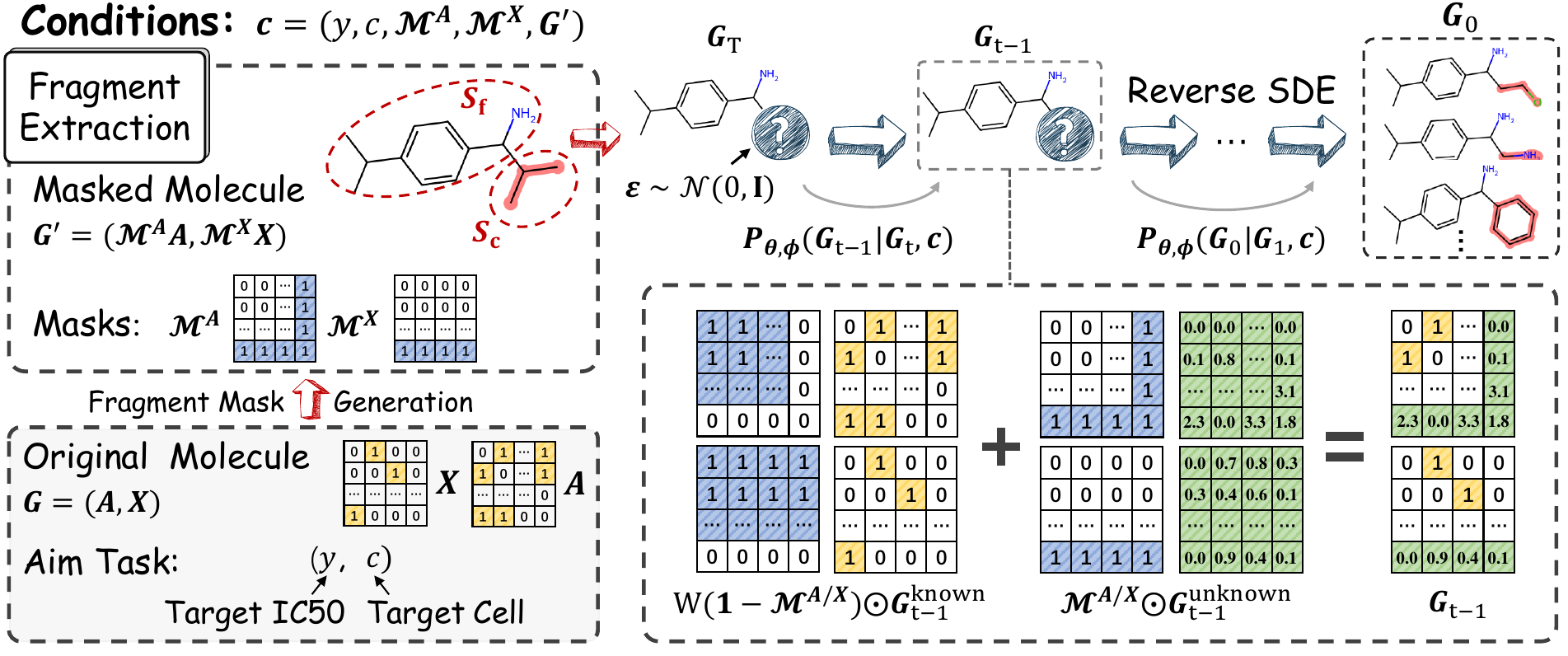}
    \caption{Our method's framework. Our optimization method involves input conditions, including one molecule to be optimized $\mathbf{G}$ and the target conditions $\mathbf{c}$. Specifically, the target conditions include an IC50 value $y$ and one cell line $c$. In addition, the molecule to be optimized is processed through the scaffold $\mathcal{S}_{\mathbf{f} }$ to identify the regions that require optimization, generating the node  $\mathcal{M}^X$ and the adjacency matrix mask $\mathcal{M}^A$.}
    \label{fig:framework}
\end{figure*}

\section{Method}

\textbf{Problem Formulation.} Molecular optimization aims to enhance a molecule's properties to reveal improved alternatives. For the PDD-based optimization task, the molecule's property is its efficacy in a specific cell line, denoted as IC50. Table \ref{tab:notation} provides a summary of the notation used in this paper.

Let one molecule be represented as $\mathbf{G}$, the cell line as $\mathbf{C}=\left \{ c_1,c_2,...,c_m  \right \} $, and their IC50 as $\mathbf{Y}=\left \{ y_1,y_2,...,y_m  \right \}$. The optimized molecule's $\mathbf{Y}$ with respect to $\mathbf{C}$ is denoted as $\mathbf{Y}'$. Therefore, for a given $(\mathbf{G}, c_i, y_i)$, the optimized result $y'$ should satisfy $y'<y$. Based on our fragment-masked method, we decomposed $\mathbf{G} $ into a scaffold $\mathcal{S}_{\mathbf{f} }$ and a side chain $\mathcal{S}_{\mathbf{c} }$, where a mask marks the side chain's fragment structure. For one molecule $\mathbf{G} = (A, X)$,  $\mathcal{M}^X \in \mathbb{R}^{\left | X \right |}$ denotes an ordered mask matrix of the atom matrix $X \in \mathbb{R}^{\left | X \right |}$, where the atoms on the scaffold are labeled as 0 and those on the side chain as 1. In addition, $\mathcal{M}^A  \in \mathbb{R}^{\left | X \right | \times\left | X \right |}$ represents an ordered mask matrix of the edge matrix $A \in \mathbb{R}^{\left | X \right | \times\left | X \right |}$, with edges on the side chain labeled as 1 and those on the scaffold or between scaffolds and side chains labeled as 0.



\begin{table}[htpb]
\centering
\caption{Mathematical Notations}
\renewcommand\arraystretch{1.2}
\resizebox{\columnwidth}{!}{%
\begin{tabular}{cl}
\toprule
\textbf{Notations} & \textbf{Descriptions} \\ \hline
$\mathbf{G}$ & Molecule graph \\
$A$ & Edge adjacency matrix \\
$X$ & Node feature matrix \\
$\mathbf{C} = \{ c_1, c_2, ..., c_m \}$ & Set of cell lines \\
$\mathbf{Y} = \{ y_1, y_2, ..., y_m \}$ & IC50 values \\
$\mathbf{Y}' = \{ y_1^{'}, y_2^{'}, ..., y_m^{'} \}$ & Optimized IC50 values \\
$\mathcal{S}_{\mathbf{f}}, \mathcal{S}_{\mathbf{c}}$ & The scaffold and side chain from one molecule \\
$\mathbf{c}=(c_i,y_j)$ & Sampling condition \\
$\mathcal{M}^A, \mathcal{M}^X$ & Mask matrix (edge and node) \\
$A^{\text{ukn}}, X^{\text{ukn}}$ & Unknown regions mask matrix (edge and node) \\
$A^{\text{kn}}, X^{\text{kn}}$ & Known regions mask matrix (edge and node) \\ \toprule
\end{tabular}
}
\label{tab:notation}
\end{table}


 
\noindent \textbf{Overview.} To generate molecules with a specific distribution under numerical drug response conditions, we employed a regressor-free conditional diffusion method. As illustrated in Fig. \ref{fig:framework}, we integrated specific conditions about the cell line $\mathbf{C}$ and IC50 $\mathbf{Y}$ into the scoring estimation to guide the diffusion model. Specifically, to establish the molecules' conditional constraints, we split the molecule into its scaffold and side chain to generate the fragment masks. Then, we performed molecule splitting according to the Murcko scaffold method. 

Finally, during the sampling phase, the drug response and fragment mask jointly constrained the sampling process, generating specific fragments that met the conditions in the mask regions.

\subsection{Molecular Conditional Generation}
\label{MCG}


Typically, the input conditions for molecular optimization methods are categorical. To more precisely and efficiently optimize molecular graphs to specific conditional distributions, our model for conditional generation follows the regressor-free molecular generation method \cite{Regressorfree}, which can effectively generate molecules under the given numerical conditions. Only the cell line type needs to be adjusted when the optimization target changes. Various attribute conditions, such as cell line types, have been fully incorporated during the training phase, and different attributes are unified through contrastive learning using a contrastive learning strategy. Moreover, the conditions comprise text labels for the PDD task (i.e., the cell line name and IC50 values) and fragment mask arrays $\mathcal{M}^X$ and $\mathcal{M}^A$. To effectively receive text conditions, we employed a contrastive learning strategy to align the two feature types. The drug, cell line, and fusion drug response encoders are denoted as $\Phi_{\mathrm{G}}$, $\Phi_{\mathrm{C}}$, and $\Phi_{\mathrm{F}}$ respectively. 

Moreover, the text encoder that describes the reaction process between the drugs and cell lines is denoted as $\Phi_{\mathrm{T}}$, and its input text $\Phi_{\mathrm{T}}$ is generated through a template with two parameters ($\mathbf{C},\mathbf{Y}$), as \textit{the response value of the drug with the $\mathbf{\color[RGB]{63, 114, 175}{\left [ name\; of\;the\;cell\;line \right ] }}$ is $\mathbf{ {\color[RGB]{63, 114, 175}{\left [ \mathbf{IC}_{50}  \right ]}}}$}. For the $i$-th representations $(d_i,c_i)$ generated by the $\Phi_{\mathrm{F}}$ and the $j$-th captions  $(c_j,y_j)$ produced by the caption encoder in a batch $\mathcal{B}$, we normalized the feature vectors in a hyper-sphere using $u _i :=\frac{ \Phi_{\mathrm{F}}(d _i,c _i)}{\left \|  \Phi_{\mathrm{F}}(c _i) \right \| } $ and $v _j :=\frac{ \Phi_{\mathrm{T}}(c _j,y_j)}{\left \|  \Phi_{\mathrm{T}}(c _j,y_j) \right \| } $. Finally, the similarity between $u _i$ and $v _j $ was calculated as $u_i^\mathrm{T} v _j$. Hence, the supervised contrastive loss function $\mathcal{L}_{\mathrm{NCE}}$ can be expressed as:

\begin{equation}
    \label{eq:nce}
    \begin{aligned}
        \mathcal{L}_{\mathrm{NCE}} = -\frac{1}{N} \left( \sum_{i}^{N}\mathrm{log} \frac{\mathrm{exp} ( u_i^\mathrm{T} v _i/\sigma )}{  {\textstyle \sum_{j=1}^{N}} \mathrm{exp} ( u_i^\mathrm{T} v _j/\sigma )} +  
        \right.
        \\
        \phantom{=\;\;}
        \left.
        \sum_{i}^{N}\mathrm{log} \frac{\mathrm{exp} ( v_i^\mathrm{T} u_i/\sigma )}{  {\textstyle \sum_{j=1}^{N}} \mathrm{exp} (v_i^\mathrm{T} u_j /\sigma )} \right)
    \end{aligned}
\end{equation}

\noindent where, $N$ is the size of the batch $\mathcal{B}$, and $\sigma$ is the temperature for scaling the logits.


By pre-training $\Phi_\mathrm{T}$ using contrastive learning, we ensured that its encoding space is aligned with that of $\Phi_\mathrm{F}$. Subsequently, we adopted an approach similar to the classifier-free guidance method, using the pre-trained contrastive model $\Phi_\mathrm{F}$ as a conditional encoder. To guide the generation process towards the desired sampling conditioning information $\mathbf{c} = \Phi_\mathrm{T}(c_i,y_i) $, we sampled the conditional distribution $q_0(\mathbf{G}|\mathbf{c})$, and carried the expectations over to the samples $\mathbf{G}_0\sim p_{data}$ and $\mathbf{G}_t\sim p_{0t}(\mathbf{G}_t|\mathbf{G}_0,\mathbf{c})$. Thus, the transition probability $p_{0t}(\mathbf{G}_{t}|\mathbf{G}_{0},\mathbf{c})$ can be represented as follows:
\begin{equation}
    p_{0t}(\mathbf{G}_{t}|\mathbf{G}_{0},\mathbf{c}) = p_{0t}(X_{t}|X_{0},\mathbf{c})p_{0t}(A_{t}|A_{0},\mathbf{c}).
\end{equation}

For time $t$, we introduced objectives \cite{songyang} to generalize score matching and estimate the scores as follows:
\begin{equation}
    \label{eq:expectations1}
    \begin{aligned}
    \underset{\theta }{\mathrm{min} } \mathbb{E}_t\left \{ \lambda _1(t)\mathbb{E}_{\mathbf{G}_0}\mathbb{E}_{\mathbf{G}_t|\mathbf{G}_0}\left \|  B_{\theta,t }(\mathbf{G}_t,\mathbf{c})-\right.\right.
    \\
    \phantom{=\;\;\}}
    \left.\left. 
    \nabla_{X_t} \mathrm{log} \; p_{0t}\left ( X_t|X_0,\mathbf{c} \right ) \right \|^2_2   \right \}
    \end{aligned}
\end{equation}

\begin{equation}
    \label{eq:expectations2}
    \begin{aligned}
    \underset{\phi }{\mathrm{min} } \mathbb{E}_t\left \{ \lambda _2(t)\mathbb{E}_{\mathbf{G}_0}\mathbb{E}_{\mathbf{G}_t|\mathbf{G}_0}\left \|  B_{\phi,t }(\mathbf{G}_t, \mathbf{c})- 
    \right.\right.
    \\
    \phantom{=\;\;\}}
    \left.\left. 
    \nabla_{A_t} \mathrm{log} \; p_{0t}\left ( A_t|A_0,\mathbf{c} \right ) \right \|^2_2   \right \}
    \end{aligned}
\end{equation}

\noindent where, $\lambda_1(t)$ and $\lambda_2(t)$ are positive weighting functions and $B_{\phi}$ and $ B_{\theta}$ denoted the noise prediction models based on the graph neural networks (GNNs) \cite{GDSS, controlnet} to estimate scores $\nabla_{A}\mathrm{log}p_{t}(X_t, A_t, \mathbf{c})$ and $\nabla_{X}\mathrm{log}p_{t}(X_t, A_t, \mathbf{c})$, respectively. These two noise prediction models are jointly referred to as $\epsilon_\theta(\mathbf{G},\mathbf{c})$.

\subsection{Fragment Mask Generation}



Scaffolds typically refer to a molecule's core structure or main ring system that determines its basic shape and properties \cite{DeepScaffold}. In contrast, side chains are the branches or functional groups attached to the scaffolds. By altering the side chain properties, we can modulate the molecule's solubility, polarity, reactivity, and various properties. As a result, we designed a fragment-based molecular optimization method. Molecular fragmentation is primarily based on the scaffolds and side chains in the molecule's structure, and we applied rule-based constraints to the fragmentation results.


For a specific molecule $\mathbf{G}= (A, X)$, we first analyzed the molecule's scaffold using the Murcko scaffold function in the RDKit tool (denoted as $\mathcal{F}_\mathrm{MS}(\cdot)$), extracting its core scaffold structure  $\mathcal{S}_{\mathbf{f} }$  and side chains $\mathcal{S}_{\mathbf{c} }$.
\begin{equation}
\mathcal{S}_{\mathbf{c} }, \mathcal{S}_{\mathbf{f} }  = \mathcal{F}_\mathrm{check} (\mathcal{F}_\mathrm{MS} (\mathbf{G}))
\end{equation}

In this instance, these side chains are referred to as fragments. After excluding the fragments containing only single atoms (e.g., 'C', 'N', 'Cl', and 'F'), we verified the connectivity between $\mathcal{S}_{\mathbf{f} }$ and  $\mathcal{S}_{\mathbf{c} }$ using $\mathcal{F}_{check}(\cdot)$.
\begin{align}
\mathcal{F}_{\mathrm{check}} = 
\begin{cases}
1, & \text{if } |\text{Connect}(\mathcal{S}_{\mathbf{c}}, \mathcal{S}_{\mathbf{f}})| = 1 \\
0, & \text{if } |\text{Connect}(\mathcal{S}_{\mathbf{c}}, \mathcal{S}_{\mathbf{f}})| \neq 1
\end{cases}
\end{align}

If a fragment $\mathcal{S}_{\mathbf{c} }$ has multiple connection points to the scaffold $\mathcal{S}_{\mathbf{f} }$, it makes the optimization task very difficult but also destroys the original scaffold's properties. Consequently, these fragments were not considered for optimization.



Additionally, by determining whether a fragment has only one atom connected to the retained scaffold with the function $\mathcal{F}_\mathrm{check}(\cdot)$, we ensured its independent optimizability. This is because  $\mathcal{M}^X$ is generated based on  $\mathcal{M}^A$. To ensure that the information of separate chemical bonds is not disclosed, we marked the row and column elements corresponding to the atoms in the fragment as 1 in $\mathcal{M}^X$. 

Finally, the fragment $\mathcal{S}_{\mathbf{c}}$ that meets the criteria is considered for optimization. Furthermore, the fragment masking involves two matrices, used for atom and bond masking in the graph $\mathbf{G}$, respectively.
\begin{align}
\mathcal{M}^X(i) = 
\begin{cases}
1, & \text{if}\; i \in \mathcal{S}_{\mathbf{c}}  \\
0, & \text{otherwise}
\end{cases}
\end{align}

\begin{align}
\mathcal{M}^A(i, j) = 
\begin{cases}
1, & \text{otherwise} \\
0, & \text{if}\; i/j \in \mathcal{S}_{\mathbf{f}}
\end{cases}
\end{align}


Specifically, the atom indices in $\mathcal{S}_{\mathbf{c}}$ correspond to those where the elements of $\mathcal{M}^X$ are 1. In addition, the matrix $\mathcal{M}^A$ is derived from $\mathcal{M}^X$, where the elements of $\mathcal{M}^A$ are set to 0 if the atoms are part of the scaffold $\mathcal{S}_{\mathbf{f}}$.

\subsection{Fragment-Masked Molecular Optimization}



Molecular optimization aims to enhance specific molecular properties by leveraging their intrinsic information. This paper focuses on optimizing the molecules' fragment regions to improve their cell line experiment responses. Hence, we employed a trained conditional diffusion denoising model. The condition's inputs include the original molecule $\mathbf{G}$, two masks $\mathcal{M}^A$ and $\mathcal{M}^X$, and the PDD task targets $(c_i, y_i)$. The output comprises multiple structurally similar molecules $\mathbf{G}'$ that exhibit improved IC50 values, denoted as $y_{i}'$, for the given cell line $c_i$.

Specifically, as each reverse step from $\mathbf{G}_t$ to $\mathbf{G}_{t-1}$ relies solely on $\mathbf{G}_t$, it is essential to guide the masked (unknown) region generation according to the known regions of $\mathbf{G}_t$ and the input optimization targets, described as follows:
\begin{align}
    \left\{ 
    \begin{array}{l}
    A^{\text{kn}}_{t-1} \sim \mathcal{N} \left(\sqrt{\bar{\alpha}_t} A_0, (1 - \bar{\alpha}_t)\mathbf{I}\right) \\
    X^{\text{kn}}_{t-1} \sim \mathcal{N} \left(\sqrt{\bar{\alpha}_t} X_0, (1 - \bar{\alpha}_t)\mathbf{I}\right)
    \end{array} 
    \right.  \\
    \left\{ 
    \begin{array}{l}
    A^{\text{ukn}}_{t-1} \sim \mathcal{N} \left(\mu_{\theta}(A_t, t), \textstyle \sum_{\theta}(A_t, t)\right) \\
    X^{\text{ukn}}_{t-1} \sim \mathcal{N} \left(\mu_{\phi}(X_t, t), \textstyle \sum_{\phi}(X_t, t)\right)
    \end{array} 
    \right.
\end{align}

\noindent where, $A_0$ and $X_0$ are the adjacency and node matrices of the initial molecule $\mathbf{G}_t$ at time $t=0$, $\beta$ is the schedule function, and $\bar{\alpha }_t = {\textstyle \prod_{i=1}^{t}} (1 - \beta_i)$. In addition, the reverse process is modeled by two neural networks (the details can be found in Eqs. \ref{eq:expectations1} and \ref{eq:expectations2}) that predict the parameters $\mu_{\theta/\phi}(\cdot)$ and $\Sigma_{\theta/\phi}(\cdot)$ of the Gaussian distributions with the given conditions.


Finally, at time step $t-1$,  unknown $(A^{\text{ukn}}, X^{\text{ukn}})$ and  known regions $(A^{\text{kn}}, X^{\text{kn}})$ are identified, constrained using two masks, and combined to form $\mathbf{G}_{t-1}(X_{t-1},A_{t-1})$:
\begin{align}
  \left\{ 
    \begin{array}{l}
    A_{t-1} = \mathrm{W}  \mathcal{M}^{A} \odot  A^{\text{unk}}_{t-1} + (1-\mathcal{M}^{A} ) \odot A^{\text{kn}}_{t-1}\\
    X_{t-1} = \mathrm{W}  \mathcal{M}^{X} \odot X^{\text{unk}}_{t-1} + (1-\mathcal{M}^{X}) \odot X^{\text{kn}}_{t-1}
    \end{array} 
    \right.
\end{align}


\noindent where, $\odot$ denotes the element-wise product, "kn" and "ukn" are the abbreviations for "known" and "unknown," respectively. $\mathrm{W}$ is a coefficient that gradually decreases from 1 to 0 over time t, and is used to control the scaffold's influence on the sampled region. After combining the known and optimized generated regions using the masks, the resulting $\mathbf{G}_{t-1}$ is incorporated into the next denoise step as follows: 
\begin{equation}
\widetilde{\epsilon}_{\theta}(\mathbf{G}_{t-1},\mathbf{c})=w\epsilon_\theta(\mathbf{G}_{t}+\epsilon, \mathbf{c})+(1-w)\epsilon_\theta(\mathbf{G}_{t}+\epsilon,\emptyset )
\end{equation}

\noindent where, the noise $\epsilon \sim \mathcal{N}(0, I)$, $w$ is a conditional control strength parameter ($w\ge0$), and $w=0$ indicates unconditional generation.



\textbf{Rule-Based Chemical Bond Post-Processing.} During the discretization of the sampled molecular graph, discretization errors may occur with continuous edge features, which could result in the generation of chemically unreasonable or unstable structures. Furthermore, the GNN-based score prediction model's inherent limitations in the molecular generation process prevent each atom from obtaining information beyond the GNN layer's $k$-hop neighborhood. This limitation may cause the model to miss global features or long-range interactions between atoms when generating molecular structures, which can, in turn, affect the overall structural rationality and stability of the molecule \cite{postprocessing,postprocessing1}. Therefore, global optimization is necessary.  To address this issue, we employed a rule-based chemical bond post-processing method. This method automatically detects and corrects potential structural inconsistencies in the generated molecules after the initial structure generation process.


\begin{table*}[t!]
\caption{Comparison of several methods for drug design across different cell lines for the PDD task.  \textbf{Success Rate} represents the proportion of cell line types in which at least one optimized molecule is found. }
\label{tab:main_compare}
\renewcommand\arraystretch{1.2}
\tabcolsep=5pt
\resizebox{\textwidth}{!}{%
\begin{tabular}{llcccccccccc}
\hline
\multicolumn{2}{c}{\multirow{2}{*}{Methods}}  & \multicolumn{2}{c}{Cell: 906792} & \multicolumn{2}{c}{Cell: 687800} & \multicolumn{2}{c}{Cell: 684055} & \multicolumn{2}{c}{Cell: 908149} & \multicolumn{2}{c}{Total Cell (985)} \\ \cline{3-12} 
 &  & N@100  & Improv.   & N@100  & Improv.   & N@100  & Improv.   & N@100  & Improv.   & Success Rate & Improv.   \\ \hline
VAE-based 
& JT-VAE \cite{JT-VAE}   & \underline{18}     & 2.89\%    & 18     & 4.07\%    & 9      & \underline{5.69\%}    & 8      & 2.74\%    & 92.70\%      & 3.4\%      \\

Graph-based & GeoLDM \cite{GeoLDM}   & 1      & 4.30\%    & 2      & 6.50\%    & 1      & 3.88\%    & 1      & 3.46\%    & 69.40\%      & 2.6\%      \\
& DiGress \cite{DiGress} & 3      & 3.40\%    & 4      & 6.40\%    & 2      & 0.12\%    & 3      & 2.76\%    & 69.30\%      & 3.0\%      \\

Diffusion-based & GDSS \cite{GDSS}       & -      & -         & -      & -         & -      & -         & -      & -         & 0.80\%       & 1.1\%      \\
& MOOD \cite{MOOD}       & 1      & 0.70\%    & 7      & 3.80\%    & 1      & 0.51\%    & -      & -         & 34.80\%      & 1.7\%      \\
& CDGS \cite{CDGS}       & 3      & 0.80\%    & 3      & 3.90\%    & 1      & 0.01\%    & 3      & 0.61\%    & 32.50\%      & 1.6\%      \\

Fragment-based & MARS \cite{MARS}                  & 4      & 5.63\%    & 2      & \underline{10.25\% }  & 4      & 5.76\%    & 19     & 6.02\%    & \underline{93.50\% }     & 6.1\%      \\
& FFLOM \cite{FFLOM}                 & 12     & 7.06\%    & 6      & \textbf{12.97\%}   & 1      & \textbf{6.62\% }   & 16     & \underline{8.61\%}    & 88.83\%      & \underline{6.3\%}      \\
& DST \cite{DST}                   & \textbf{44}     & 2.63\%    & \underline{22}     & 4.97\%    & 7      & 5.16\%    & -      & -         & 55.53\%      & 4.1\%      \\
&  Prompt-MolOpt \cite{Prompt-MolOpt}         & 9      & \underline{7.42\% }   & 2      & 1.36\%    & 2      & 1.36\%    & 3      & 6.88\%    & 91.68\%      & 5.7\%      \\
& HN-GFN  \cite{HN-GFN}               & 12     & 3.55\%    & 10     & 5.99\%    & 8      & 5.74\%    & 14     & 3.74\%    & 92.70\%      & 3.3\%      \\ \hline

  \multicolumn{2}{c}{FMOP (Ours)}                     & 15     & \textbf{7.80\%}    & \textbf{62}     & 5.57\%    & \textbf{67}     & 4.58\%    & \textbf{85}     & \textbf{9.53\% }   & \textbf{95.43\%}      & \textbf{7.5\%}      \\ \hline
\end{tabular}%
}
\end{table*}

\begin{figure*}[t!]
\centering
\subfloat[\footnotesize For the cell line 906792]{\includegraphics[width=1.75in]{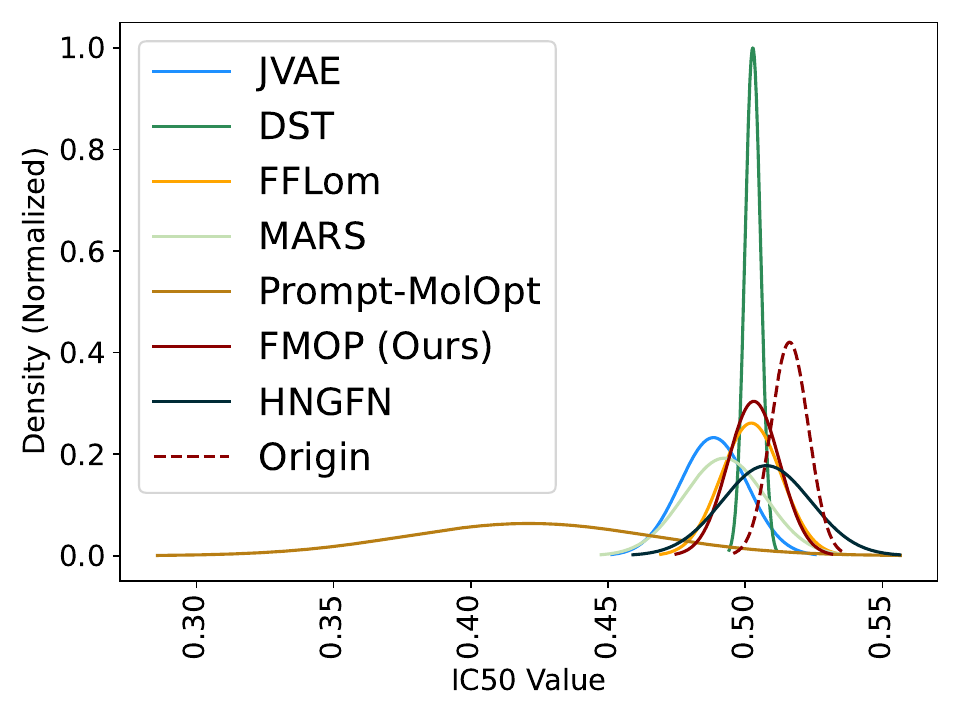}%
\label{fig:dis_image1_2}}
\hfil
\subfloat[\footnotesize For the cell line 687800]{\includegraphics[width=1.75in]{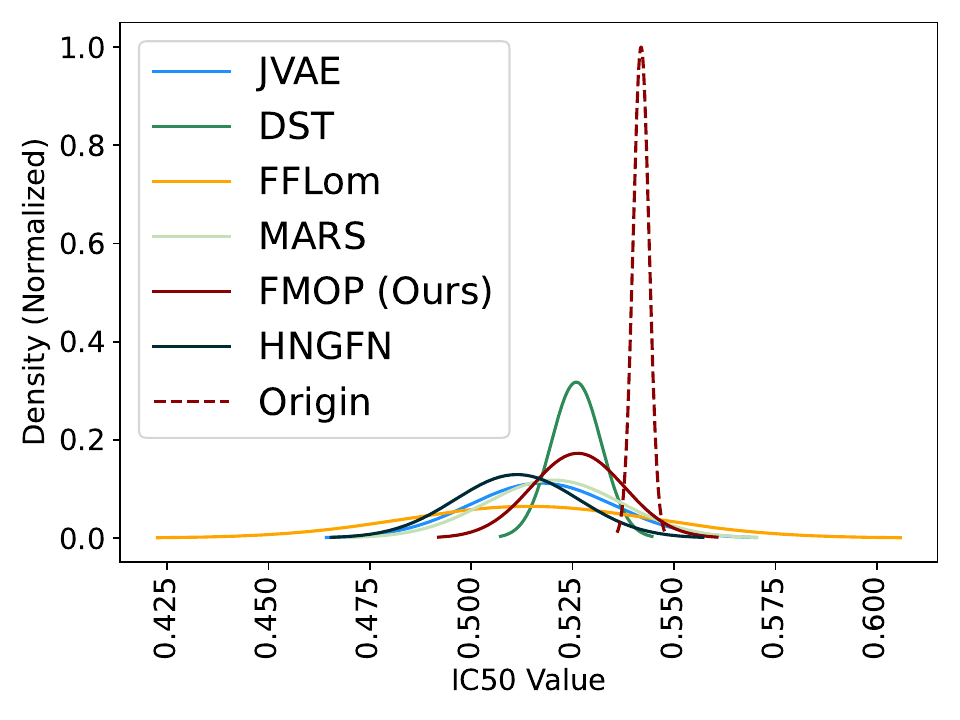}%
\label{fig:dis_image2_2}}
\hfil
\subfloat[\footnotesize For the cell line 684055]{\includegraphics[width=1.75in]{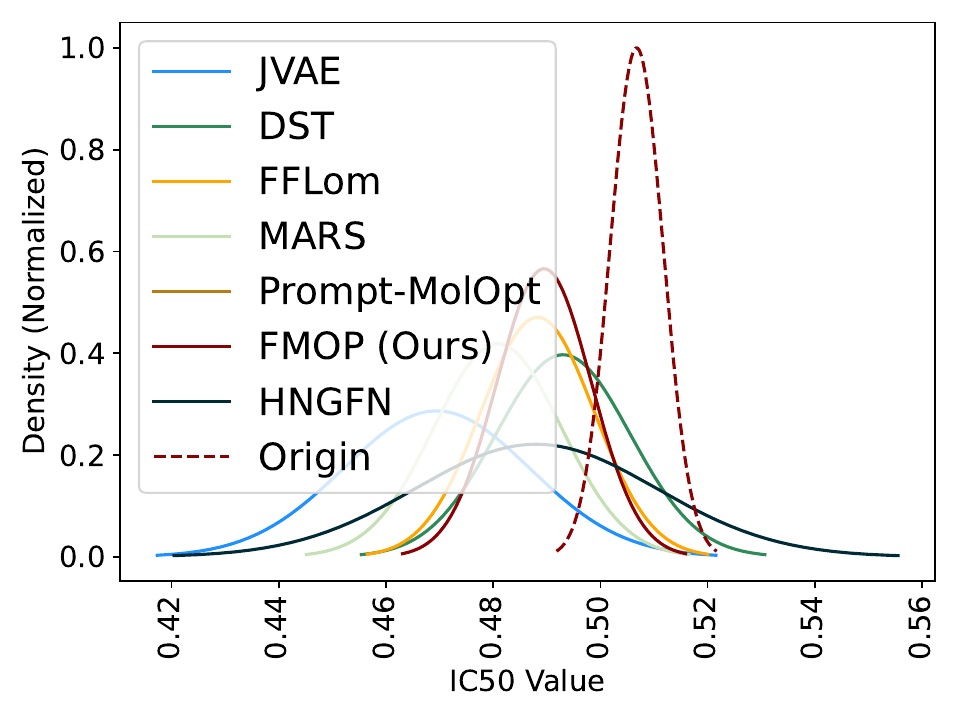}%
\label{fig:dis_image3_2}}
\hfil
\subfloat[\footnotesize For the cell line 908149]{\includegraphics[width=1.75in]{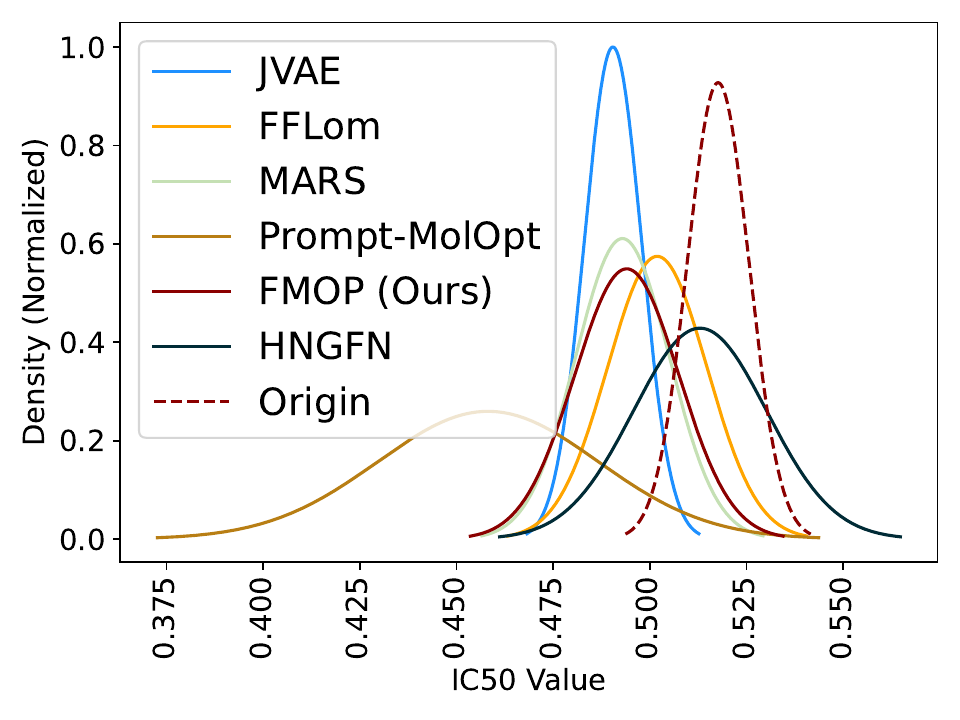}%
\label{fig:dis_image4_2}}
\caption{Visualizations results for the IC50 distribution of molecules generated by fragment-based methods.}
\label{fig:combined_2}
\end{figure*}

\begin{figure*}[t!]
\centering
\subfloat[\footnotesize For the cell line 906792]{\includegraphics[width=1.75in]{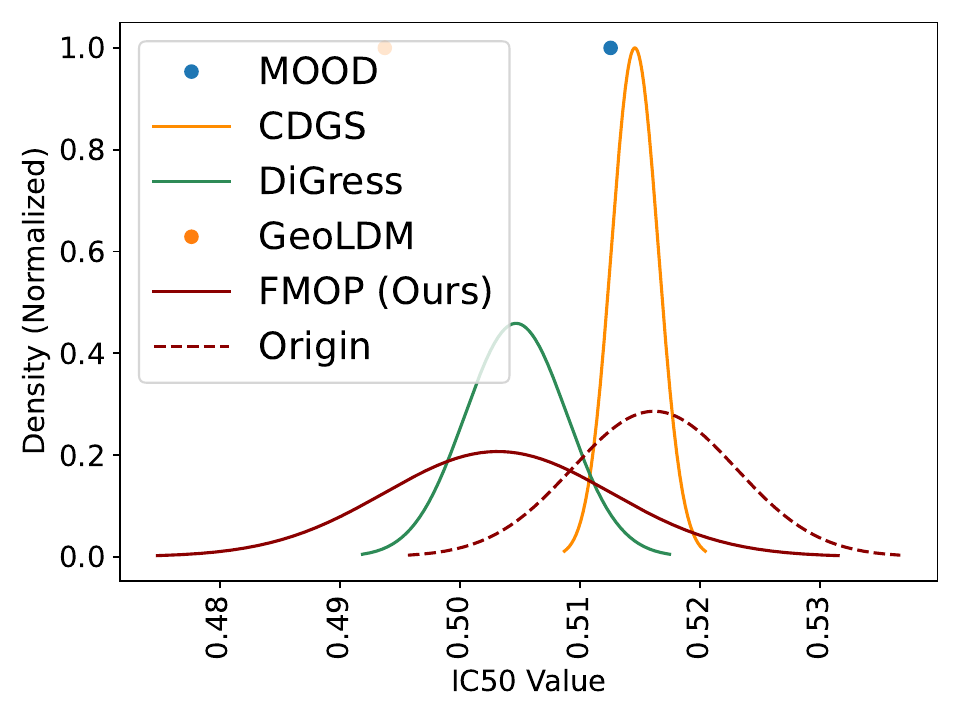}%
\label{fig:dis_image1_1}}
\hfil
\subfloat[\footnotesize For the cell line 687800]{\includegraphics[width=1.75in]{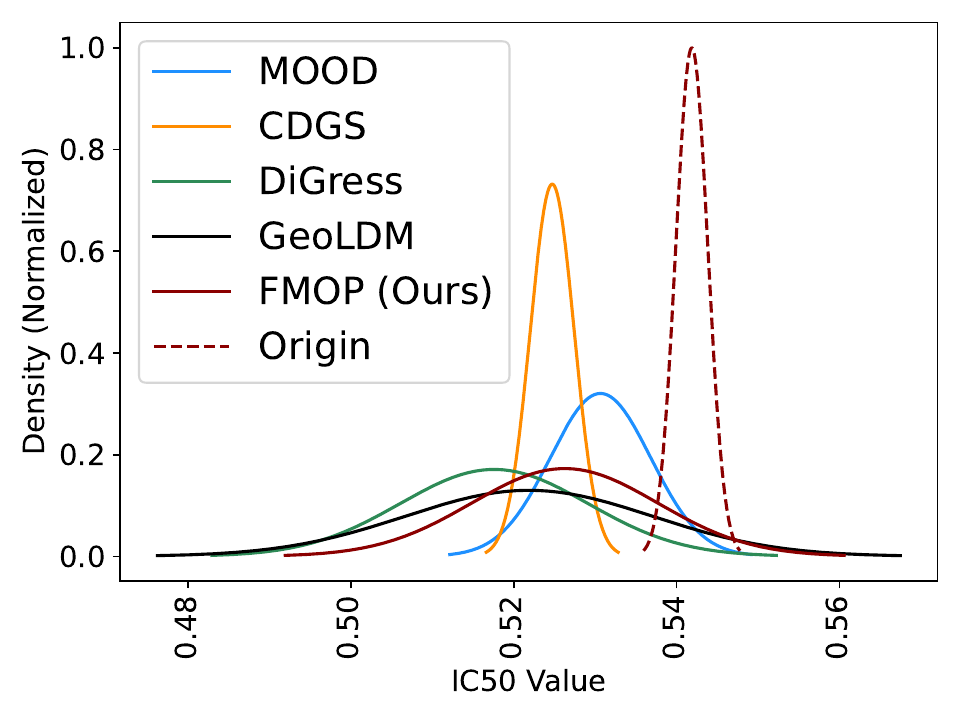}%
\label{fig:dis_image2_1}}
\hfil
\subfloat[\footnotesize For the cell line 684055]{\includegraphics[width=1.75in]{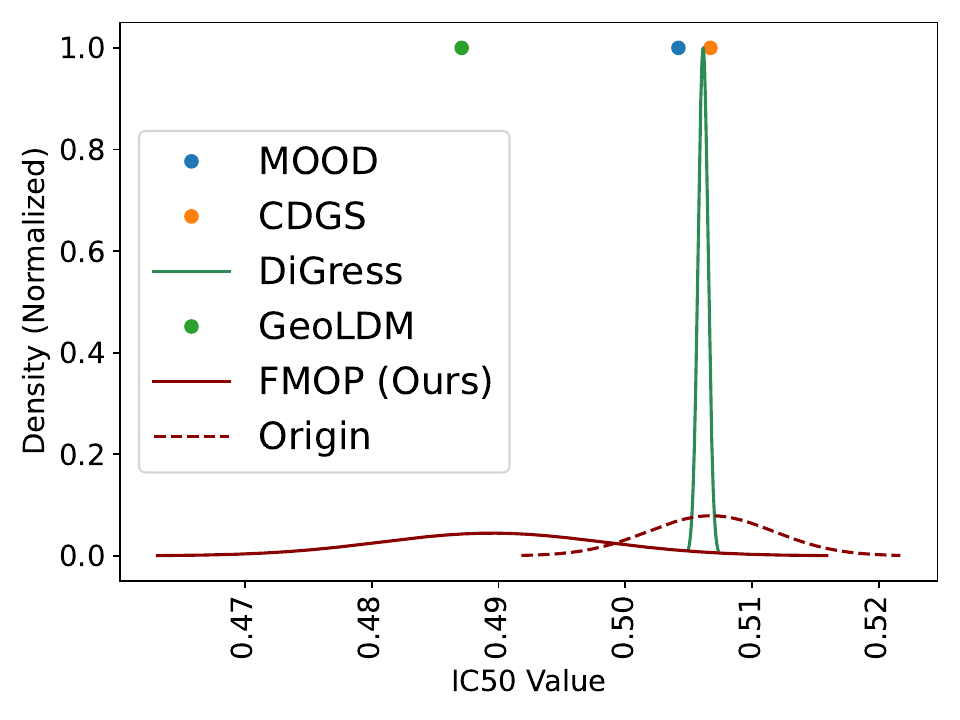}
\label{fig:dis_image3_1}}
\hfil
\subfloat[\footnotesize For the cell line 908149]{\includegraphics[width=1.75in]{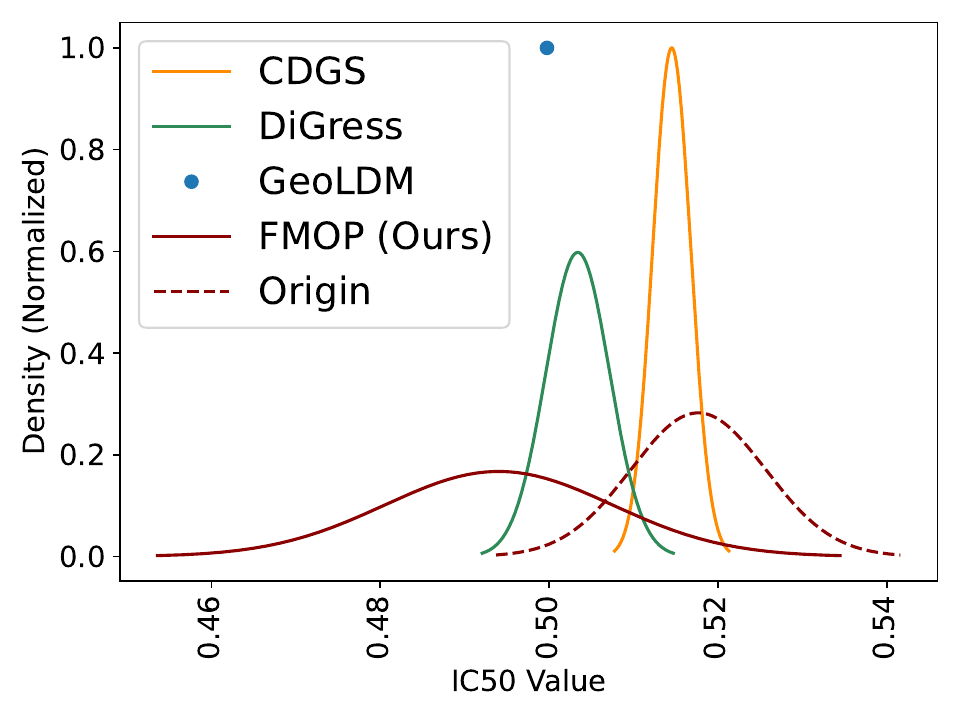}%
\label{fig:dis_image4_1}}
\caption{Visualizations results for the IC50 distribution of molecules generated by graph- and diffusion-based methods.}
\label{fig:combined_1}
\end{figure*}

Specifically, we modified the molecular structure in the following ways:
\begin{enumerate}
    \item Conversion of continuous double bonds to single bonds: This change is intended to prevent the formation of unstable chemical structures. In certain chemical reactions, consecutive double bonds can give rise to reactive intermediates that are both unstable and highly reactive.
    \item Modification of six-carbon chains with double bonds into aromatic rings: This transformation improves stability, as aromatic rings are generally more stable than alkenes due to their conjugated electron structures.
\end{enumerate}

Through this post-processing step, we not only improve the chemical stability of the generated molecules but also ensure their structural rationality. The rule-based method integrates chemical knowledge with computational models, resulting in molecules that better conform to actual chemical principles and drug development requirements, thereby enhancing the accuracy and effectiveness of molecular optimization.


\section{Experiment}

\subsection{Experimental Setup}



This study utilized two primary datasets: QM9 \cite{QM9} and GDSCv2 \cite{GDSC}. The QM9 dataset was used for pre-training the model to enhance molecular diversity, and contains approximately 133,885 molecules. These data provide the model with rich molecular information, improving its generalization ability. Likewise, the GDSCv2 dataset was used for tasks related to drug response prediction \cite{drp3,drp1}, and comprises approximately 190,853 samples, covering 985 cell lines and 220 drugs. GDSCv2 enables learning and predicting drug response distributions. This is significant for precision medicine and new drug development \cite{drp2}.


%

\textbf{Evaluation Criteria.} Molecules labeled with drug response for the cell line $\eta$ are selected from the GDSCv2 dataset, focusing on those with IC50 values in the top 20\% to 30\%. These molecules, referred to as $y_{ \eta }$, were used as the target molecules for optimization. \textbf{N@100} is a counting function, if the IC50 of the optimized molecules  $y_{\eta}^{'}$ is lower than the average $y_{\eta}$ and the reduction exceeds 1\%, the count is incremented by one. The \textbf{Improv.} represented the improvement in IC50 before and after molecular optimization. Our method calculated the average increase based on the true IC50 values for each molecule being optimized, while other methods also used the average IC50 of these molecules to compute the average increase. \textbf{Success Rate} $ =  \frac{ {\textstyle \sum_{i=1}^{\mathrm{M}}} \left \{ \mathrm{N} _i@100>0  \;\mathrm{and}\; \mathrm{Improv}(i)>1\%   \right \} }{\mathrm{M}} $ represents the proportion of  985 cell line types (denoted as $\mathrm{M}$) in which at least one optimized molecule is found.

It is important to note that the generated or optimized molecules' the efficacy (i.e., the IC50) in various cell lines was predicted using the deep learning models and not wet lab validation. This is because the wet lab validation involves molecular synthesis and cell-based assays, processes that are time-consuming and extremely costly. Therefore, we relied on the out-of-domain drug response prediction (OOD-DRP) methods \cite{MSDA, OOD-DRP2, OOD-DRP1} to predict the IC50 of the newly generated molecules. When drug information is unavailable during the training phase and only cell line types are used for training, the testing accuracy typically ranges from the Pearson correlation coefficient \cite{OOD-DRP1} of 0.6 to 0.8. This accuracy depends on factors such as the distribution of the dataset, the diversity of cell lines, and the complexity of the predictive model used. The contrastive learning drug response model based on natural language supervision (CLDR) focused on out-of-domain generalization and demonstrated state-of-the-art correlation in zero-shot response predictions \cite{CLDR}. Consequently, we used the CLDR as the OOD-DRP method.

\subsection{Baselines}

To validate the effectiveness of FMOP, this study has meticulously ensured fairness by comparing it with eleven baseline methods:

\begin{itemize}
    \item JT-VAE \cite{JT-VAE}: A variational autoencoder (VAE) model \cite{VAE} for molecular graph generation. It first generates a tree-structured scaffold library, and then combines selected samples from the scaffold library into the molecules using the graph message-passing network \cite{GMPN}.
    
    \item GeoLDM \cite{GeoLDM}: A latent diffusion model designed for 3D molecular generation, using autoencoders to encode structures into latent codes and diffusion models to operate in the latent space. 
    
    \item DiGress  \cite{DiGress}: A discrete denoising diffusion model that iteratively adds or removes edges and modifies categories. A graph transformer network is trained to reverse this process.
    
    \item GDSS  \cite{GDSS}: A graph generative model based on score-based diffusion, utilizing a system of stochastic differential equations (SDEs) \cite{SDEs} to model the joint distribution of molecular nodes and edges. It generates molecules that adhere to chemical valency rules and closely follow the training distribution.

    \item MOOD  \cite{MOOD}: A score-based diffusion model for exploring chemical space, utilizing out-of-distribution (OOD) \cite{OOD_data} control in the generative process to generate novel molecules. It is conditioned on target properties such as drug-likeness and synthesizability, guiding the diffusion process toward high-quality molecules.

    \item CDGS  \cite{CDGS}: A conditional diffusion model for molecular graph generation, incorporating OOD control in a generative SDEs to explore novel regions of chemical space.
    
    \item MARS  \cite{MARS}: A multi-objective drug discovery method that iteratively edits molecular graph fragments using Markov Chain Monte Carlo (MCMC) \cite{MCMC} sampling and the GNNs.
    
    \item FFLOM  \cite{FFLOM}: A flow-based autoregressive model for fragment-to-lead optimization, which generates molecular structures by linking fragments and growing R-groups.
    
    \item DST  \cite{DST}: A differentiable scaffolding tree method for molecular optimization, which converts discrete chemical structures into locally differentiable ones for gradient-based optimization.
    
    \item Prompt-MolOpt  \cite{Prompt-MolOpt}: A molecular optimization method that leverages prompt-based embeddings to enhance the transformer's ability to optimize molecules for targeted properties.
    
    
    \item HN-GFN  \cite{HN-GFN}: A multi-objective Bayesian optimization (MOBO) \cite{MOBO} algorithm that uses a hypernetwork-based GFlowNets \cite{GFlowNets} (HN-GFN) as an acquisition function optimizer.
    

\end{itemize}



\begin{table*}[t!]
\caption{The table presents the optimized molecular properties of 12 molecular optimization methods, including molecular weight (MW, typically ranging from $>50$g/mol to $<500$g/mol), the log of the partition coefficient of a solute between octanol and water(LogP, typically ranging from $-1.5$ to $<5$), hydrogen bond donors (HBD, typically ranging from $0$ to $5$), hydrogen bond acceptors (HBA, typically ranging from $1$ to $10$), polar surface area (PSA, typically ranging from $<20 \ \text{Å}^2$ to $>140 \ \text{Å}^2$), rotatable bonds (ROTB, typically ranging from $0$ to $15$), and quantitative estimation of drug-likeness (QED, typically ranging from $0$ to $1$).}
\label{tab:Mol}
\tabcolsep=18pt
\renewcommand\arraystretch{1.1}
\resizebox{\textwidth}{!}{%
\begin{tabular}{lccccccc}
\hline
Methods & MW & LogP & HBD & HBA & PSA & ROTB & QED \\ \hline
Base & 452.25 & 3.6175 & 2.2332 & 6.4484 & 98.9591 & 5.7399 & 0.4641 \\ \hline
JVAE & 305.47 & 2.5448 & 0.9769 & 3.7962 & 60.2302 & 4.5326 &  0.8134 \\
MOOD & 283.39 & \cellcolor[HTML]{FFC7CE}{\color[HTML]{9C0006} 5.6476} & \cellcolor[HTML]{FFC7CE}{\color[HTML]{9C0006} 0.0469} & \cellcolor[HTML]{FFC7CE}{\color[HTML]{9C0006} 0.3971} & \cellcolor[HTML]{FFC7CE}{\color[HTML]{9C0006} 5.2514} & 4.3908 & 0.5848 \\
GDSS &  128.17 & 0.2786 & 1.9167 & 2.0833 & 53.2942 & 3.1667 & 0.5702 \\
CDGS &  122.20 & 0.7358 & 1.3031 & 2.0334 & 41.0165 & 2.8278 & 0.5386 \\
DiGress &  123.47 & 0.4331 & 0.6862 & 2.6745 & 33.4768 & 1.1279 & 0.5241 \\
GeoLDM &  113.80 & 0.3699 & 1.8906 & 2.0741 & 51.4864 & 1.0894 & 0.5103 \\
MARS & 404.88 & 3.2676 & 1.7030 & 5.8044 & 84.4286 & 5.6015 & 0.4476 \\
Prompt-MolOpt & \cellcolor[HTML]{FFC7CE}{\color[HTML]{9C0006} 516.12} & 3.2727 & 2.6674 & 8.0261 & 126.2293 & 6.3037 & 0.3500 \\
DST & 495.81 & 4.8649 & 2.4808 & 6.7039 & 101.4287 & 7.4716 & 0.3407 \\
FFLom & \cellcolor[HTML]{FFC7CE}{\color[HTML]{9C0006} 524.99} & 4.3199 & 1.8279 & 7.1002 & 99.6118 & 8.0676 & 0.3322 \\
HNGFN & \cellcolor[HTML]{FFC7CE}{\color[HTML]{9C0006} 533.64} & 3.4179 & 3.0587 & 8.7698 & \cellcolor[HTML]{FFC7CE}{\color[HTML]{9C0006} 155.8988} & 5.0059 & 0.3322 \\
FMOP (Ours) & 439.41 & 3.1507 & 3.2866 & 6.2079 & 96.8763 & 6.6725 & 0.3567 \\ \hline
\end{tabular}%
}
\end{table*}

\begin{table}[t!]
\caption{The ablation study shows the impact of different components on success rate, improvement, and total N@100 (TN@100).}
\centering
\label{tab:ablation}
\renewcommand\arraystretch{1.1}
\resizebox{\columnwidth}{!}{%
\begin{tabular}{cccc}
\hline
Methods  & Success Rate & Improv. & TN@100  \\ \hline
w/o. Fragment Mask & 0.3\%      & 2.1\%     & 26           \\
w/o. Task Guidance  & 5.0\%      & 2.1\%     & 1278          \\
w/o. Modification  & 70.8\%      & \textbf{9.4}\%     & 12352          \\ \hline
Origin Method    & \textbf{95.4}\%     & 7.5\%     & \textbf{23789}       \\ \hline
\end{tabular}
}
\end{table}

\subsection{Overall Experiments}

To verify whether our method can effectively optimize molecules to achieve better drug response values, we conducted overall experiments involving various diffusion models. Due to the novelty of the molecule generation method, generated molecules are out-of-domain and require the OOD-DRP model to have a high generalization capability. Therefore, we used the CLDR method \cite{CLDR}, which has excellent generalization performance, to predict the optimized molecules. 


It is crucial to highlight that the optimization task involving 985 cell lines is similar to optimizing across various properties. Our approach enables a single training process to address all tasks simultaneously, whereas other methods would require separate training for each of the 985 cell lines individually. Table \ref{tab:main_compare} displays the optimization results of 12 methods in 4 different cell line scenarios and the average optimization results across 985 cell lines. Our method achieves the best optimization and increase rate results. 

\begin{figure*}[ht!]
    \centering    \includegraphics[width=0.999\linewidth]{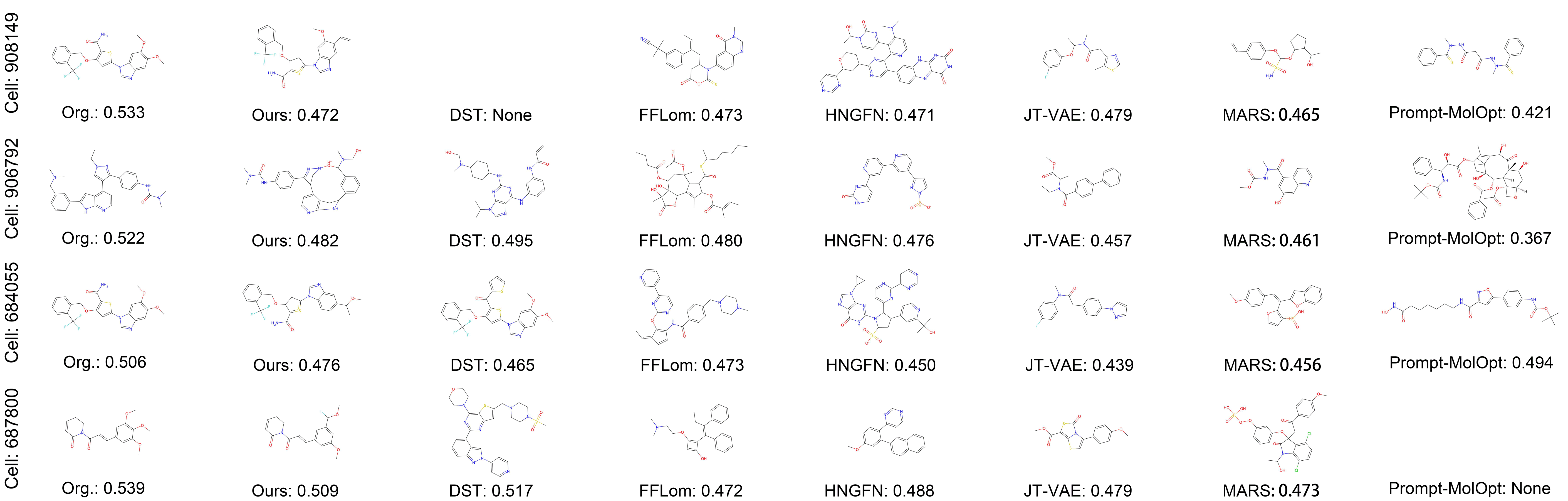}
    \caption{Visual comparison of our optimization method with generative methods. This illustrates the unique molecular structures generated by our method and compares them with various baselines across four distinct cell lines. Our method consistently produces diverse and effective molecules tailored to each cell line, avoiding convergence to the same local optimum.}
    \label{fig:mol_compare}
\end{figure*}

In order to confirm that our method can effectively optimize fragment regions while maintaining scaffold consistency, we conducted a visual comparison of the results. As shown in Fig. \ref{fig:mol_compare}, our method maintains scaffold consistency while optimizing the masked region. Based on the distribution of cell line attribute features, FMOP employs regressor-free guidance, effectively generating molecules with specific attributes. The scaffold structures of these molecules are similar to the original ones being optimized, with the property values improved through side-chain optimization. In contrast, molecules generated by other optimization methods exhibit significant differences from the original molecules. Even fragment-based methods (DST, JT-VAE, and Prompt-MolOpt) still show poor performance. This is because these fragment-based methods essentially rely on statistical fragments from the training set to form a fragment library. During the molecular optimization phase, specific fragments are conditionally selected from the library and stitched together, but these methods cannot optimize fragments within the scaffold. 

Some graph- and diffusion-based methods (GDSS, CDGS, DiGress, and GeoLDM) achieved success (1$\sim$3\%) mainly by randomly generating a few molecules with good efficacy across various cell lines, which failed to adjust the sampling space distribution for specific tasks, resulting in suboptimal performance. To this end, we analyzed the molecular properties, such as MW, logP, QED, etc. In early drug design, the Lipinski’s rule of five \cite{rof}, is commonly used as a guideline for evaluating drug candidates. We have marked results that fall outside the normal property ranges in red. As shown in Table \ref{tab:Mol}, the average MW and LogP of the molecules generated by these methods are significantly lower than those in the original dataset. Specifically, the original dataset has an average MW of approximately 452, while the molecules generated by these methods have an average MW of about 120. This indicates that the generated molecules have notably smaller molecular weights, which could potentially lead to adverse changes in their pharmaceutical properties.






\subsection{Ablation Study}

\begin{figure}[t!]
    \centering
    \includegraphics[width=0.99\linewidth]{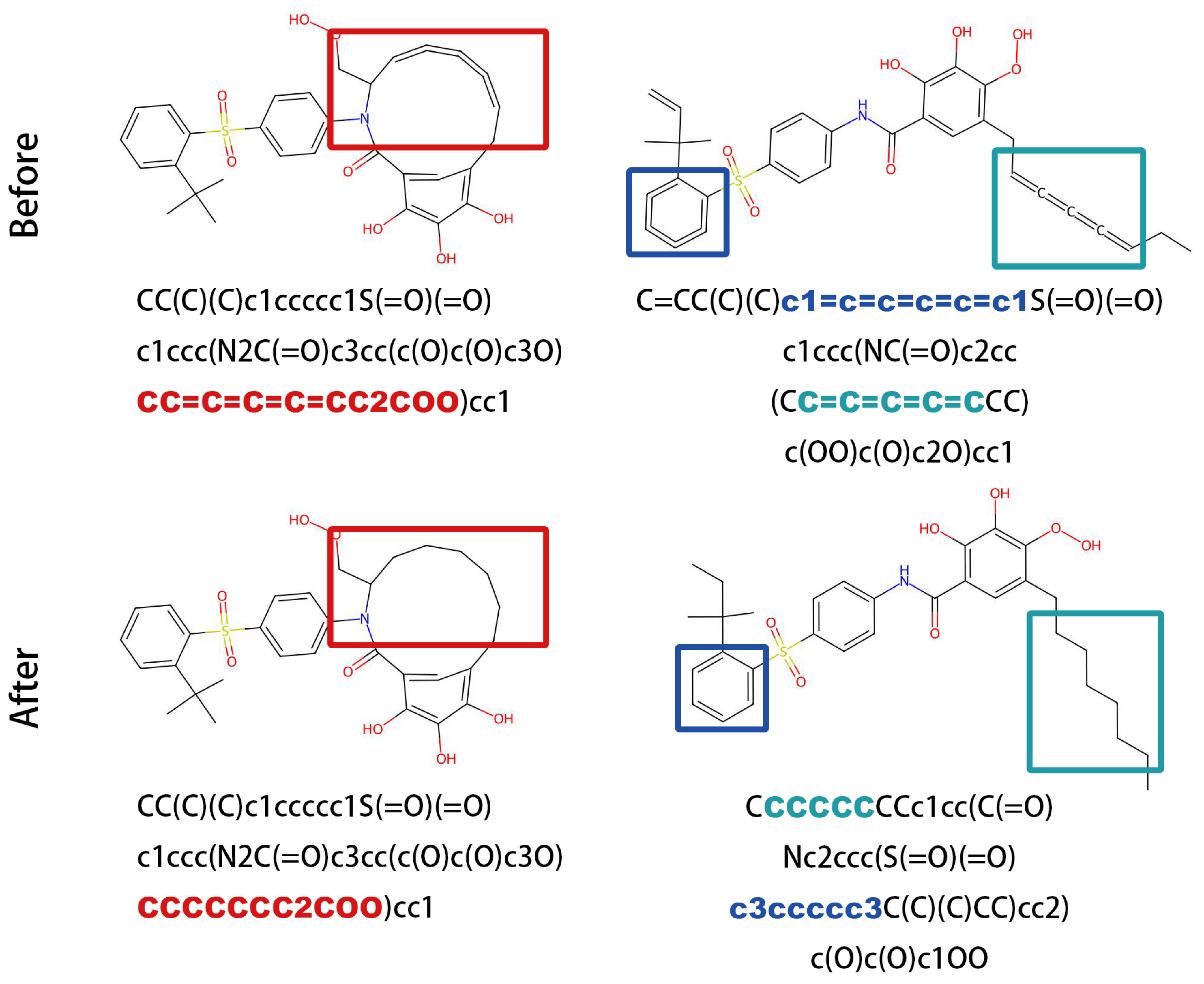}
    \caption{Visualization results from rule-based chemical bond post-processing.}
    \label{fig:correct}
\end{figure}

In the FMOP method, the PDD task information and fragment masks are encoded as indispensable conditions, and their impact on the final optimization results is significant. Therefore, three key points need to be explored:

\textbf{Q1}: Do the conditional information (i.e., the expected IC50 values and cell line types) play a crucial role in the molecular optimization process, thereby improving optimization success rates and efficacy enhancements?

\textbf{Q2}: Can fragment masking effectively focus optimization on specific regions to improve efficacy and optimization success rates?

\textbf{Q3}: Given the current issues with aromatic ring quality and single/double bond generation, is rule-based chemical bond post-processing an effective method for molecular generation?

\begin{figure*}[t!]
    \centering
    \includegraphics[width=0.92\linewidth]{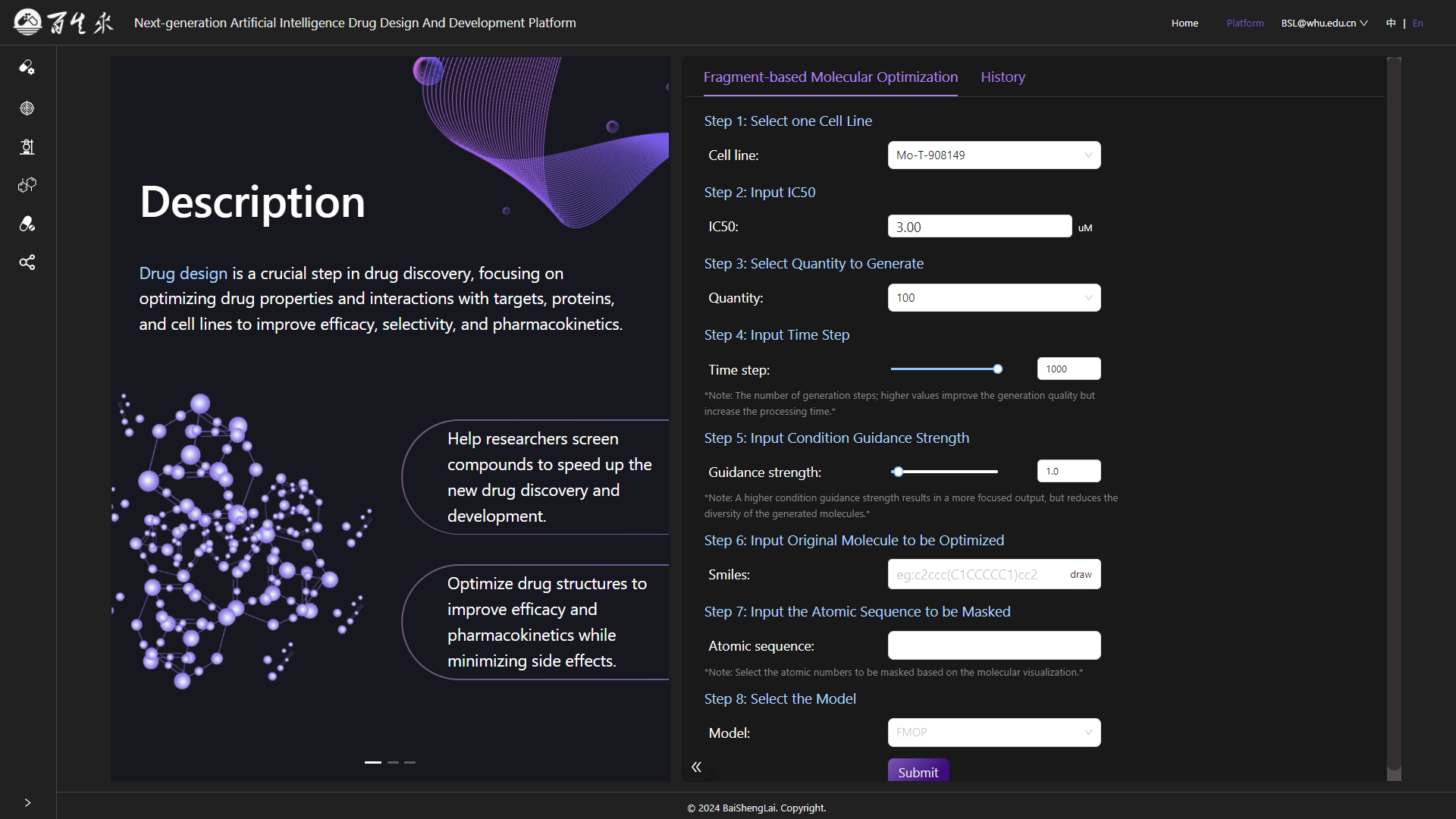}
    \caption{Front-end visualization of the FMOP method on the Beishenglai platform. Users input the target conditions and submit the optimization task.}
    \label{fig:bsl_overview}
\end{figure*}


As shown in Table \ref{tab:ablation}, each component significantly contributes to the model's overall performance. For Q1, without the fragment mask prompt, our method generates molecules randomly, resulting in a drastic decline in the success rate to 0.3\% and an improvement of 2.1\%, with only 26 instances reaching N@100. This indicates that the fragment mask is crucial for identifying key molecular features.

For Q2, when task guidance is removed, the success rate is 5.0\%, and the improvement rate remains at 2.1\%. The absence of task guidance leads to \textbf{a random fragment generation} strategy. In the QM9 dataset, there are 1798 fragments with a frequency of occurrence greater than 10, which we have collected into a fragment library. For each optimization task, we randomly select 100 fragments from this library. These fragments are then attached to the atom in the original molecule that has the most implicit hydrogen atoms. The atom types are limited to 'C', 'N', 'O', 'S', and 'P'. The first atom of each selected fragment is connected to the target atom via a single bond to prevent the covalent bond from exceeding the threshold set for the central atom, resulting in a success rate of 5.0\%.  This demonstrates that task guidance is essential for potential molecule optimization. 



For Q3, the success rate further increased to 95.4\% through post-processing based on the aromatic ring recognition rules. This suggests that molecular modifications are essential for achieving a high success rate, although they introduce some complexity.

Overall, these findings highlight the importance of fragment masks, task guidance, and post-processing rules in enhancing the quality of molecular generation, particularly in improving the generation of aromatic rings.

\subsection{Visualization Analysis}


To explore whether the molecules generated or optimized using different methods achieved a certain confidence level instead of merely evaluating the methods based on numerical values, we visually analyzed the molecular structures generated using our optimization method and baselines across different cell lines. As shown in Fig. \ref{fig:mol_compare}, our method generates unique molecules for each cell line, ensuring that the optimization process does not converge to the same local optimum across different cell lines. In comparison, the molecules generated using other methods were generally similar, highlighting that our method is able to optimize molecular structures based on the specific response values of each cell line, thereby achieving more effective and suitable molecular structures. Furthermore, our method's predicted IC50 values remain consistently low across different cell lines, indicating that our optimized molecules have a competitive advantage compared to de novo designed molecules.

Since measuring the IC50 for all virtually generated molecules on 985 specific cell lines in a short time is impractical, we utilized the CLDR method to predict these values. To validate our method's effectiveness, as shown in Fig. \ref{fig:combined_1} and Fig. \ref{fig:combined_2}, we predicted the IC50 values for the cell lines 906792, 687800, 684055 and 908149, then visualized the mean and variance by assuming a Gaussian distribution. As a result, our proposed FMOP method demonstrated a strong competitiveness in the lower IC50 range. For example, as shown in Fig. \ref{fig:combined_1}(c), the IC50 values for the molecules generated by FMOP range from 0.46 to 0.52, whereas the molecules generated by other methods typically include only one successfully optimized molecule (represented by $\bullet$). Additionally, the IC50 distribution of the original molecules (denoted as 'Origin') ranged between 0.51 and 0.52. 



\begin{figure}[t!]
    \centering
    \includegraphics[width=0.99\linewidth]{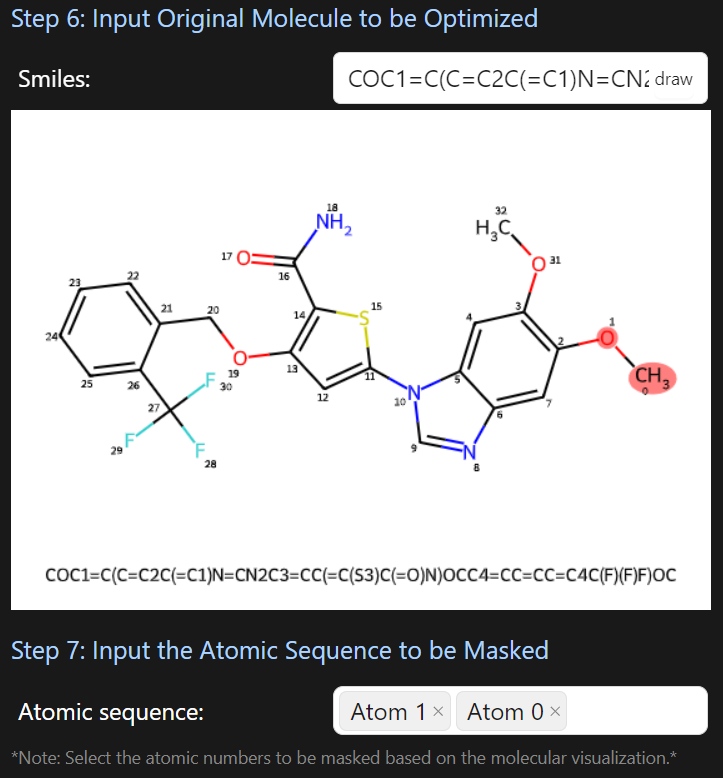}
    \caption{Detailed interface for steps 6 and 7, where users interactively select specific atomic positions within a molecule for fragment optimization. The selected fragment must contain at least one atom, and the optimization region is not limited to specific fragment libraries or functional groups.}
    \label{fig:bsl_detail}
\end{figure}

\subsection{Case Study}

We have deployed our method on the Beishenglai platform \footnote{The online platform can be accessed at \url{https://www.baishenglai.com}.}, a drug discovery platform based on deep learning models. The platform supports key drug discovery processes: generation, optimization, prediction, and retrosynthesis. To showcase the practical application of our method, we present a case study. In this case study, our FMOP method was employed to optimize the efficacy of the molecule Z-LLNle-CHO (Compound ID: 16760646, uniquely identified in the PubChem database) \footnote{The SMILES representation of Z-LLNle-CHO is CCCCC(C=O)NC(=O) [C@H](CC(C)C)NC(=O)[C@@H](CC(C)C)NC(=O)OCC1=CC=CC=C1.} for a specific cancer type, Mo-T (ID: 908149 in the GDSCv2), while preserving the similarity of its molecular scaffold structure. The complete interface on the Beishenglai platform is shown in Fig. \ref{fig:bsl_overview}, where users input their target conditions and submit the optimization task. 

\begin{figure}[t!]
    \centering
    \includegraphics[width=0.99\linewidth]{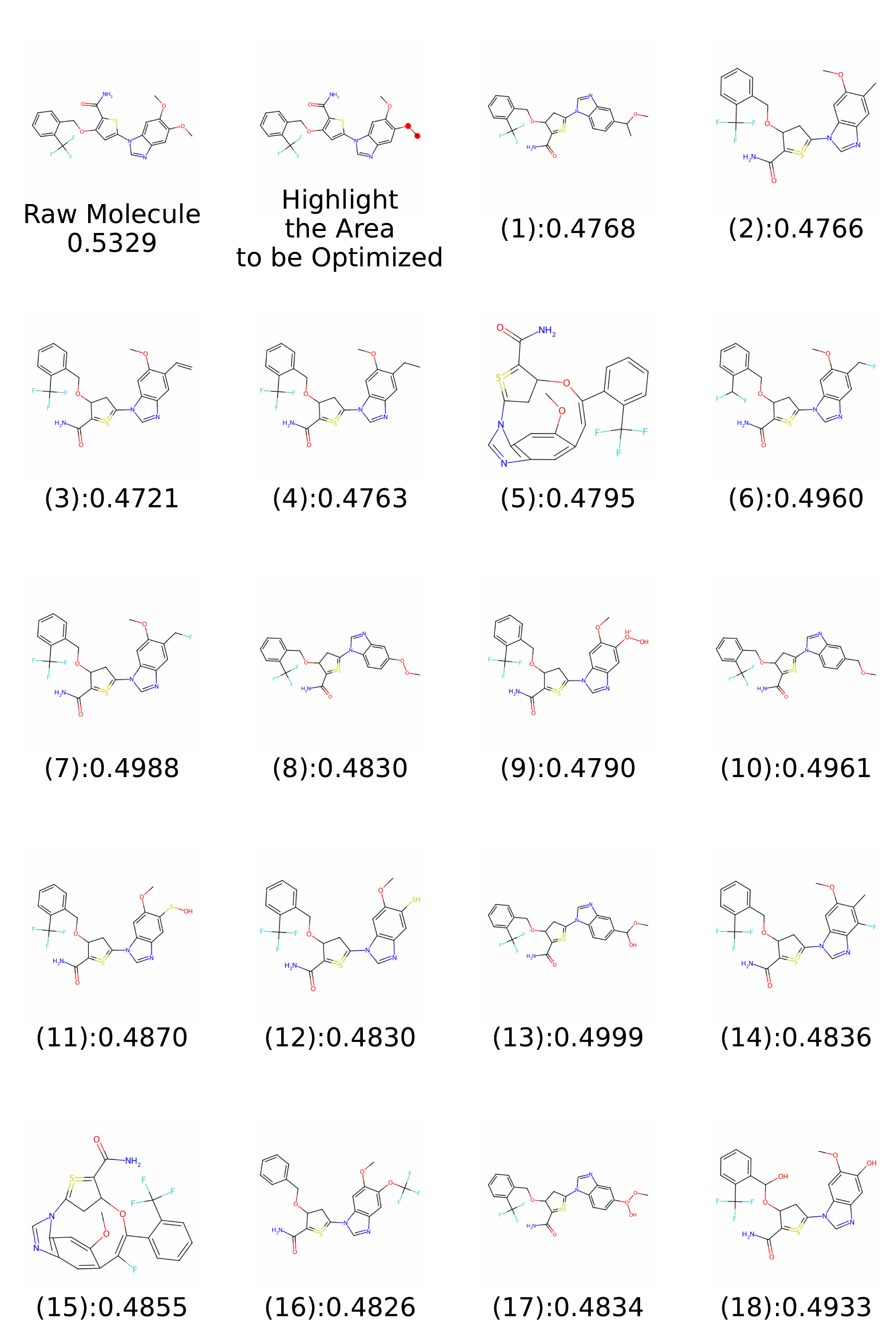}
    \caption{Visualization of case study results, with normalized IC50 values annotated below each molecule.}
    \label{fig:bsl_detail_result}
\end{figure}

The optimization process involves selecting the cell line, specifying the target IC50, choosing the number of molecules to generate, setting the diffusion time step, defining the guidance strength, and providing the original molecule. The number of molecules should be chosen based on the desired diversity, as a larger quantity may increase processing time. The time step controls optimization detail, with smaller values requiring more time. The guidance strength determines the optimization focus; higher values lead to more focused outputs but reduce diversity.

Fig. \ref{fig:bsl_detail} illustrates the details for steps 6 and 7, allowing users to select atomic positions for fragment-masked optimization. The mask must include at least one atom, and the optimization region is not restricted to specific fragment libraries.

After task submission, the FMOP method was applied, and the results are shown in Fig. \ref{fig:bsl_detail_result}. Among the 100 optimized molecules,18 had an IC50 superior to the original, with the best-performing molecule (3) achieving an IC50 of 0.4721, an 11.4\% improvement in efficacy. These results demonstrate the FMOP method’s ability to enhance key molecular properties and offer a flexible framework for optimization.





\section{Conclusion}



To address the PDD challenge of molecular optimization, which requires screening a vast number of possible molecular structures, we proposed the FMOP method. To the best of our knowledge, the FMOP is the first optimization method for the PDD task. FMOP employs a regression-free diffusion model to conditionally sample the masked regions of molecules for optimization, effectively generating new molecules with similar scaffolds and improved efficacy. We optimized the molecules for all 985 cell lines on the  GDSCv2. The overall experiments demonstrated that the in-silico optimization success rate reaches 95.4\%, with an average efficacy increase of 7.5\%. Additionally, we conducted extensive ablation studies and visualization experiments, proving that FMOP is an effective and robust molecular optimization method.

Although the FMOP method has demonstrated exceptional performance in enhancing molecular efficacy, enabling optimization across multiple task conditions with a single training session, it still has limitations. First, our method is primarily suited for optimizing molecules with existing efficacy, assuming the target molecule exhibits some initial activity. In cases where efficacy is unclear or absent, the applicability of FMOP is limited. Second, FMOP relies on fragment masking for localized optimization, which may neglect the global structural requirements of the molecule.

\clearpage

\bibliography{main}
\bibliographystyle{IEEEtran}

\end{document}